# Coupling Effects in Multi-Stage Laser Wake-field Acceleration of Electrons

Zhan Jin[1], Hirotaka Nakamura[2], Naveen Pathak[3], Yasuo Sakai[3], Alexei Zhidkov[3], Keiichi Sueda[1], Ryosuke Kodama[2,4], Tomonao Hosokai[1,3,*]

[1] Laser Acceleration Development Team, Innovative Light Sources Division, RIKEN SPring-8 Center, 1-1-1, Kouto, Sayo-cho, Sayo-gun, Hyogo, 679-5148, Japan
[2] Graduate School of Engineering, Osaka University, 2-1 Yamada-oka, Suita, Osaka, 565-0871, Japan
[3] Institute of Scientific and Industrial Research, Osaka University, 8-1 Mihogaoka, Ibaraki, Osaka, 567-0047, Japan
[4] Institute of Laser Engineering, Osaka University, 2-1 Yamada-oka, Suita, Osaka, 565-0871, Japan
[*] hosokai@sanken.osaka-u.ac.jp

**ABSTRACT**

Staging laser wake-field acceleration is considered as a necessary technique for developing full-optical jitter-free electron accelerators. Splitting of the acceleration length into several technical parts with their lengths smaller than the dephasing length and with independent laser drivers allows generation of stable, reproducible acceleration fields. Temporal and spatial coupling of pre-accelerated electron bunches for their injection in the acceleration phase of a successive laser pulse wake field is the key part of the staging laser-driven acceleration. Here, characterization of the coupling is performed with dense, stable, a narrow energy band <3% and energy selectable electron beams with charges ~1.6 pC and energy ~10 MeV generated from a laser plasma cathode. Cumulative focusing of electron bunches in a low density pre-plasma, exhibiting the Budker- Bennett effect, is shown to result in the efficient injection of electrons even with a long distance between the injector and the booster in the laser pulse wake. Measured characteristics of electron beams modified by the booster wake field agree well with those obtained by multidimensional particle-in-cell simulations.

**Introduction**

Laser wake field acceleration (LWFA) of electrons is one of the rapidly developed scientific fields for the last decade [1-16]. This technique, providing potentially jitter-free sources of radiation and electrons, has already demonstrated the electron acceleration over 4 GeV [13] in a single stage with laser pulse energy less than 100 J. However, similar to vacuum acceleration schemes, the staging schemes in plasma (an injector, a buster, and so on) seem to be more practical providing better stability and reproducibility of the acceleration process [17, 18].

Coupling in the staged schemes for LWFA is apparently a key problem. Origin of coupling problem is from a possible mismatch in sizes of injected electron bunches and sizes of laser wake fields. The size of a laser wake field is limited by values of $a_0$ (the normalized vector potential of the laser field $a_0=eE_L/mc\omega$, where $E_L$ is the laser electric field strength, $\omega$ is the laser pulse frequency [19]) necessary for the efficient acceleration, which imply an upper limit on the value of laser pulse waists, $w_0$. On the other hand, longitudinal size of the acceleration field is determined by plasma electron density: the smaller electron density, the longer wake wave. However, the laser pulse guiding requires the electron density $N_e > 1.7 \times 10^{10}\ N_{cr}[cm^{-3}]/P[W]$, where $N_{cr}$ is the critical density for the laser pulse frequency, $P$ is the total power of the laser pulse. Moreover, a too low electron density results in a weaker acceleration field $E=a_0 \lambda/\lambda_p$ where $\lambda$ is the laser wavelength and $\lambda_p =\lambda(N_{cr}/N_e)^{1/2}$ is the plasma wavelength. The sizes of electron bunches coming out of an injector are determined by (i) geometrical emittance and (ii) the energy spread $\Delta\gamma$ ($\gamma$ is the relativistic factor $\gamma =[1+(p/mc)^2]$). If a distance between an injector and a booster is $L$, the bunch length at the entrance point will be equal to $l=L\Delta\gamma/\gamma_0^3$ where $\gamma_0$ is the mean energy of electrons in the bunch. [We assume $\Delta\gamma \ll \gamma_0$ and $l$ is bigger than the initial bunch length]. In case of $l>\lambda_p$ an essential portion of electron cannot be injected in the acceleration phase of laser wake field.

It is clear also that if the bunch transverse size exceeds the laser pulse diameter in the focus spot, again an essential part of injected electron cannot be accelerated. For example, the injection efficiency of an electron bunch with its diameter ~1 mm into a wake generated by a laser pulse with its focus spot ~20 $\mu$m should be only 0.04%, almost zero. Fortunately, such a coupling cannot be simply estimated with use of only geometrical emittance of bunches. An actual efficiency may be quite high due to the Budker-Bennett effect [20, 21] in plasma. Injection of high energy electron beams in plasma results in evacuation of some plasma electrons from the beam axis owing to the beam longitudinal electric field. This process is a beginning of formation of beam wake field [22]. Electrons in beams are heavier $\gamma_0$ times than plasma electrons and cannot be evacuated. Therefore, beam electrons propagate over a positively charged part of plasma. To estimate effect of such plasma on beam electrons we use the Poison equation in the reference frame moving with beam electrons to exclude the magnetic field. One can easily find that the electric field strength obeys the following equation:

$$\frac{1}{r}\frac{\partial}{\partial r}\left(rE_{tr}\right) = -4\pi e\left[N_B / \gamma_0 - \gamma_0(N_i - N_e)\right], \qquad (1)$$

where $E_{tr}$ is the electric field strength in the transverse direction, $N_B$ is the beam density, $N_i$ and $N_e$ are the ion and electron density in plasma. If the value in the square brackets is positive, the beam electrons will move towards the beam axis or will be focused. For that the difference of the density should be $(N_i-N_e)>N_B/\gamma_0^2$. For a ball beam with 10 $\mu$m diameter and its charge ~10 pC the beam density is $N_B$~6×10$^{16}$ cm$^{-3}$. For $\gamma_0$=20 it gives $\Delta N$~10$^{14}$cm$^{-3}$. Such a condition does not require dense plasma, and beam focusing may occur even in quite low density plasma part in front of a plasma target.

In the present work, we investigate these temporal and spatial coupling effects in a booster irradiated by 450 mJ laser pulses with use of well-determined electron bunches generated from a laser-plasma cathode [23] with its charge ~1.6 pC and energy ~10 MeV with the energy spread $\Delta E$<3% (Fig. 1). To understand details of effects we also perform multidimensional particle-in-cell simulations both for the temporal coupling and for the electron focusing in various plasmas.

## Results

First of all, we characterized the injection electron bunches similarly to Ref. [23]. Fig. 2a-c gives the spectral image of the injector electron beam on the ESM measured without the function of solenoid, with 0.8 kV voltage applied on solenoid, and same voltage with an additional 500 $\mu$m diameter, 5 mm thickness molybdenum aperture at the electron focus spot, respectively. Due to the short $f$-number parabolic mirror we used for the cathode, electron beams with the thermal-like spectrum were observed, with their maximal energy up to 25 MeV, as shown in Fig. 2d. Such a setup allows us to collect the specified energy of electrons a predetermined focus point by only tuning the solenoid voltage. With the applied voltage equal 0.8 kV accelerated electrons with energy ~10 MeV were selected and focused with an energy spread around 3%. The focus spot size of the electrons is measured to be <0.7 mm FWHM with a total charge of ~1.6 pC [24-26], as shown in Fig. 2d. The simple estimation of the injection efficiency would give ~0.1% for the chosen parameters.

To separate the spectra modified upon interaction of injected electron beams with the booster plasma, first, we characterized electron spectra without the coupling. Fig. 3a and Fig.3b show the electron spectra measured for the cathode (injector) only and for the booster (dark current, without a cathode beam) only, respectively. One can see that the dark current in this experiment was negligibly small. This was due to a rather low intensity of the second laser beam. An efficient electron self-injection in the booster plasma was absent and no high energy electron beam was observed. It was reached also upon careful tuning of the gas jet position.

Modulated spectra of accelerated electrons after passing the booster are shown in Fig.3 c-f depending on the delay time. The zero time equals the time when both the injector electron beam and booster laser beam were delivered to the booster synchronously. Clear deceleration and/or acceleration of the beam electrons were observed depending on the time delay exhibiting the coupling of the injection electron beams with the laser wake field in the booster. Visible instability of the measured spectra mainly came from the temporal jitter (small but notable in the present set-up) and the vibration of the laser focus pointing. Noted that the diameter of the injector beam focused only by the solenoid is hundreds of micrometers, which is much larger than the booster wake-field transverse size. However, the results show that most of the electrons can be modulated by the booster laser wake-field, which indicates that there is possibly a self-focusing by the beam induced plasma wake-field, when the injector electrons propagating in the low density gas region before the main jet.

Spectrum behavior strongly depends on the geometrical parameters of injected beam at the position of the booster, $L$~1.4 m. We measured the transverse size of the injected beams as 0.7 mm without the booster plasma. Such a big diameter is a result of geometrical emittance of cathode beams. While the beam length $l=L\Delta\gamma/\gamma_0^3$ should be (for $\Delta\gamma/\gamma_0$~2-3% and $\gamma_0$=20) equal 70-100 $\mu$m. The spot size of the second laser beam is about $D$~20 $\mu$m and the length of the laser wake field is about $\lambda_p$~10 $\mu$m for parameters of the experiment. It seemed to be difficult to observe an essential coupling for such circumstances. However, the coupling was observed. Moreover, the number of counts for accelerated electrons with modified energy after passing through the booster varied from 10% to 90% from the initial. These values exceed the expected in orders of magnitude! We attribute this to the effect of Bennet-Budker [20, 21]. According to the measurement result [27], the gas jets used in the experiment have a long, several mm front part with relatively low gas density $N$~$10^{17}$ cm$^{-3}$. In the booster this part of gas flux becomes plasma after the second laser beam pass through due to optical field ionization.

To verify this effect, we performed multidimensional particle-in-cell simulations for electron beams propagating in the low density plasma $N_e$=3×$10^{17}$ cm$^{-3}$. The beam diameter was 50 $\mu$m, electron energy $E$=10 MeV, the energy spread $\Delta E/E$=3%, the total charge 10 pC, and geometrical emittance was $10^{-2}$ rad. We considered three plasma target: uniform plasma, convex and concave density plasma channels. The ratio the minimal electron density to the maximal in the case of channels is equal 0.7. In Fig. 4 the longitudinal and transverse beam profiles are given after 30 ps propagation time in plasma. One can see an essential focusing of the electron bunches in all kind of plasma. The weakest focusing is observed for a concave-plasma channel, while the strongest for a convex-plasma channel as seen in Fig.4d and Fig. 4f. This reflects a simple fact: evacuation of plasma electrons runs easier for the negative density gradients. This proves that long pre-plasma in front of the booster can focus much stronger the injected electron beams to diameter enough for the efficient coupling of these electrons with the booster wake field. We anticipate that this process may become a critical in developing of beam lines for the dense, high charge electron beams generated by laser wake fields.

However, the beam length cannot be shortened by plasma, and it remains for 10 MeV-electron bunches with a few percentage energy spread longer than the length of wake wave. Therefore, accelerated electrons were injected in the different phases of the laser wake field in the booster. Final energy distribution of the electron should have accelerated, decelerated, and intact parts. To reveal the process of interaction of well-determined electron beam with the laser wake field we perform multidimensional particle-in-cell simulation with extraction of beam electrons from plasma electron in post-processing.

Typical spatial distributions of accelerated electrons along with the electric field of laser wake, produced by a laser pulse with duration 30 fs and $a_0$~2, are given in Fig.5a (1D projection) and in Fig.5b (3D plot) for an electron bunch with $E$=10 MeV and its length 70 $\mu$m, which essentially increase the plasma wavelength. One can see a very strong modulation of electrons in the wake field. Expectedly the length of the bunch exceeds the length of wake wave. Along with the classical theory of oscillations the most of electrons are concentrated near the zero field points. Therefore it is difficult to reach an efficient acceleration or deceleration for the present conditions. A change of time delay cannot help much since the beam length far exceeds the wake wavelength.

Dynamics of the electron energy distribution is given in Fig.6 up to 4.3 ps. It confirms that only small amount of electrons are further accelerated and, correspondingly, a small amount of electron are decelerated. Most of electrons have near the same energy as before the interaction. It goes well with the experimental results given in Fig. 3 and proves the strong coupling of injected electron with the booster. We present the results of calculation for two density values. Both $N_e$=3x$10^{19}$ cm$^{-3}$ and $N_e$=3x$10^{18}$ cm$^{-3}$ give similar results for the evolution of energy distributions of 10 MeV electron bunch with 3% energy spread. However we have to note that the lower density distribution is closer to the measured.

**Discussion**

In conclusion, we have demonstrated the strong coupling of electrons accelerated with the laser pulse from a gas-jet, with wave breaking electron self-injection and the conventional beam energy slice techniques [23], and the booster produced in the second gas jet by another laser pulse temporally and spatially synchronized with the first laser beam. First, the electron beams with the energy 10 MeV charges ~1.6 pC were focused and transported 1.4 m downstream with energy spread ~3% and spatial size of less than 800 $\mu$m in FWHM, respectively. These beams were injected in the wake field generated in the second gas jet by another, synchronized laser beam.

The injected efficiency has been measured as unexpectedly high, from 10% to 90% of the initial electron charge despite of the low focus diameter of the second laser beam, $D$~10 $\mu$m. Such high efficiency is the result of Bennet–Budker effect of electron self-focusing in rather low density plasma. This has been proved by multidimensional particle-in-cell simulations with various plasmas. It has been shown that this effect may be used to organized future beam lines for laser driven dense electron beams.

The longitudinal spread of electron bunch at the distance to the booster position was too long ~80-100 $\mu$m to provide efficient electron acceleration in the wake wave with its size $\lambda_p$~10 $\mu$m. Most of electrons from the cathode are concentrated near zero fields that prevent both efficient acceleration and an efficient deceleration. The maximal energy of electrons has reached ~30 MeV with the maximal gain ~3. The results of multidimensional particle-in-cell simulations agree well with observations and have explained the dynamics of interaction of the injected electrons and laser wake field for present experimental set-up.

## Methods

**Laser System.** The experiment was carried out with the P-cube 80-TW Ti:Sapphire laser system (Amplitude Technologies) at Osaka University based on a chirped pulse amplification (CPA) technique [28]. The laser system can deliver 2 synchronized CPA laser beams from the same oscillator. The first laser beam (shown as Laser 1 in Fig.1) serves for the plasma cathode. The pulse energy on a target was $E_{L1}$=500 mJ at its maximum and the pulse duration was $\tau$=30 fs. The booster laser beam 2 had $E_{L2}$=350 mJ and 30 fs duration. The central wavelength of the laser pulse is $\lambda$=800 nm. The contrast ratio between the main pulse and the nanosecond pre-pulse caused by the amplified spontaneous emission (ASE) was set to be about $10^{-9}$. Picosecond contrast was of the order of $10^{-4}$-$10^{-5}$.

**Experiment.** The experimental setup is schematically shown in Fig.1. The first laser was focused on the front edge of the slit nozzle of a He gas-jet by a gold-coated off-axis parabolic mirror (OAP) with f/3.5 (*f* = 163 mm). The length of gas-jet was 1 mm. The stagnation pressure of the gas jet was set to 0.5-4 MPa with the gas density at the laser axis orders of $N$=$10^{18}$-$10^{19}$ cm$^{-3}$. Spatial distribution of the ejected electron beams was controlled by a phosphor screen (Mitsubishi Chemical Co. LTD, DRZ-High). The DRZ-High screen is sensitive to high-energy particles and radiations, so that the front side of the screen was laminated with a 12 $\mu$m-thick aluminum foil to avoid exposure to the laser pulses, scattering lights and low-energy electrons. The scintillating images on these phosphor screens made by the deposited electrons were recorded by charged coupled device (CCD) cameras (Bitran Co., BU-51LN) with a commercial photographic lens from the backside of the screen.

To make a well-determined beam from the plasma cathode we used a pulse solenoid lens system to extract electrons with low energy band similar to Ref. [26]. Parameters of the solenoid lens were chosen to provide necessary focus-ability for electron beams with energy from 10MeV. The solenoid was put 100mm after the injector plasma to collect and focus the electrons with desired energy to the booster plasma wake field at 1 meter away.

The booster laser beam 2 was focused by an f/20 OAP to another 1 mm size gas jet with an intensity of 5×$10^{18}$ W/cm$^2$. A mirror with a 5 mm diameter hole at the center was installed in the electron beam line to reflect the booster laser beam, while allowing the electrons from the cathode to pass through, as given in Fig.1. As well as for the cathode, the stagnation pressure of the second gas jet, the booster, was set to 1-4 MPa with the gas density at the laser axis orders of $N$<$10^{19}$ cm$^{-3}$. The temporal synchronization between these two laser beams were adjusted with the optical interference method, with a jitter measured to be lesser that 20 fs. An electron spectrometer (ESM) was installed just after the second gas jet.

**Particle-in-cell simulation.** To explore interaction of bunch electrons with the booster we used numerical simulations with 3D particle-in-cell code FPlaser3D [29] with moving window technique. In the post-processing electrons in the bunches were separated from plasma electrons to make beam dynamics clearer and better understood. For that, the 'beam' electrons were specially marked. This allowed us to explore self-consistently and directly the beam dynamics and its relaxation. The calculations performed with a spatial resolution $\lambda/30$ in the direction of laser pulse propagation and $\lambda/10$ in the transverse directions, and the temporal resolution $\Delta t=\Delta x/2$. We used 25 particles per a cell; the size of the simulation box was chosen to be big in both transverse and longitudinal directions: 200x200x150 $\mu m^3$. This allows us to explore dynamics of practical, large size beams propagating over cm scale length plasma.

## Acknowledgements

This work was supported and funded by the ImPACT Program of Council for Science, Technology and Innovation (Cabinet Office, Government of Japan). Part of this work was also supported by the JST-MIRAI Program Grant No. JPMJMI17A1. The computational work is performed by using Mini-K computing system at the SALCA facility, SPring-8.

## Competing interests

The author(s) declare no competing interests.

## Author contributions statement

Z.J., Y.S., N.P., K.S., T.H. conducted the experiments, T.H. designed the experiments, R.K., A.Z. and T.H. conceived the experiments, A.Z., H.N. and N.P. conducted the simulations. All authors reviewed the manuscript.

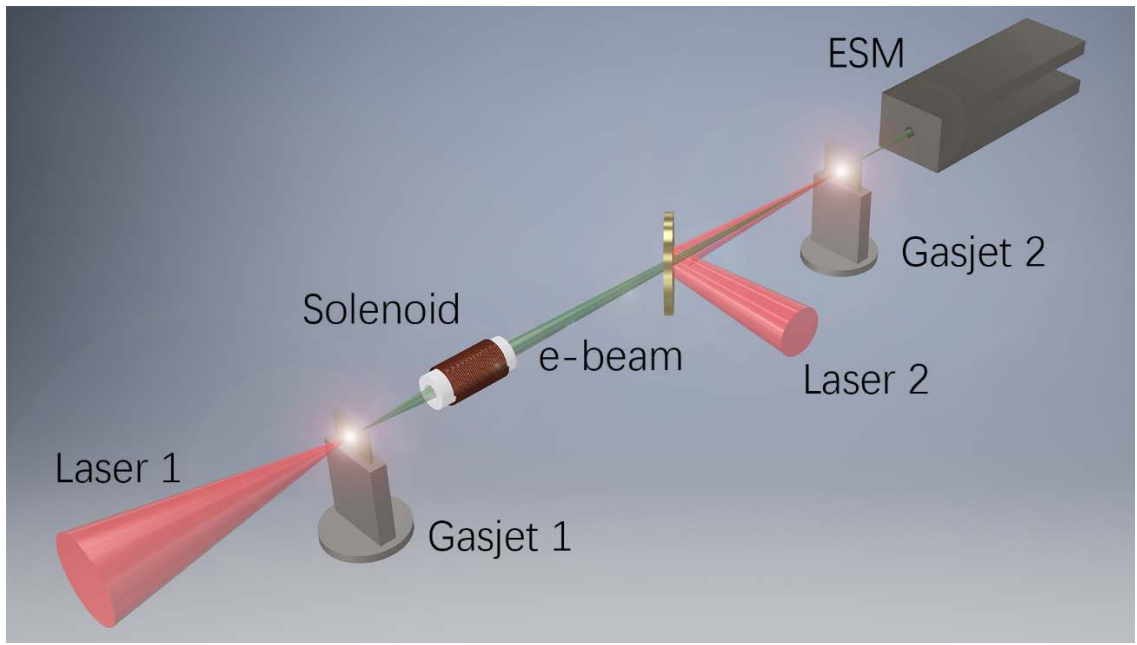

**Figure 1.** Experimental setup. The laser beam 1 was focused a He gas jet producing injector electrons. A solenoid collected and focus the electrons with desired energy to the booster plasma wake-field at 1 meter away. An electron spectrometer was installed just after the second gas jet.

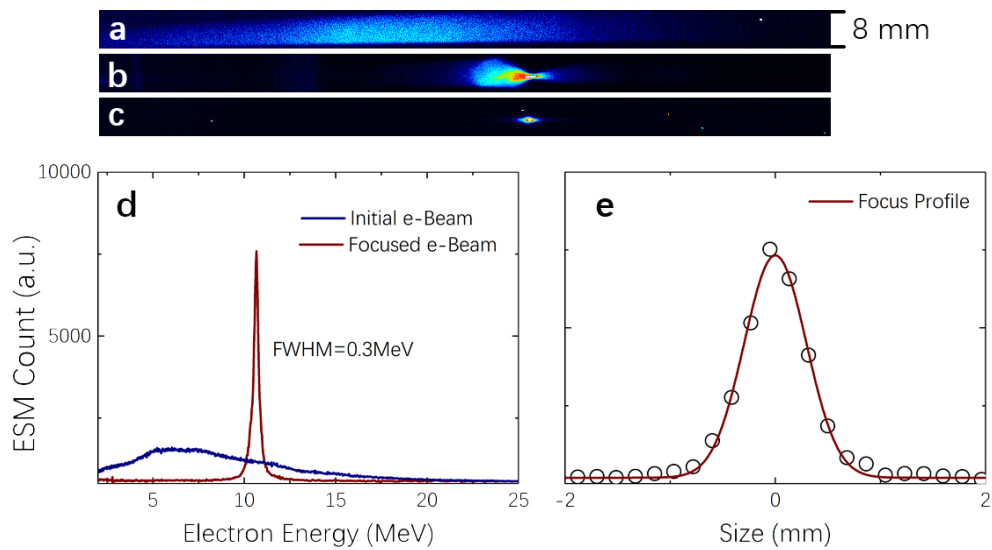

**Figure 2.** Characteristics of electron beams form the cathode: (a) Initial injector electron beam's spectrum. (b) Spectrum with 0.8 kV voltage applied on solenoid. (c, d) Spectrum with same voltage and an additional 500 $\mu$m diameter, 5 mm thickness molybdenum aperture at the electron focus spot. (e) The spatial profile of the electron's focus. Electrons with beam energy ~10 MeV can be focused with the energy spread around 3%. The focus spot size of the electrons is measured to be <0.7 mm FWHM with a total charge of ~1.6 pC.

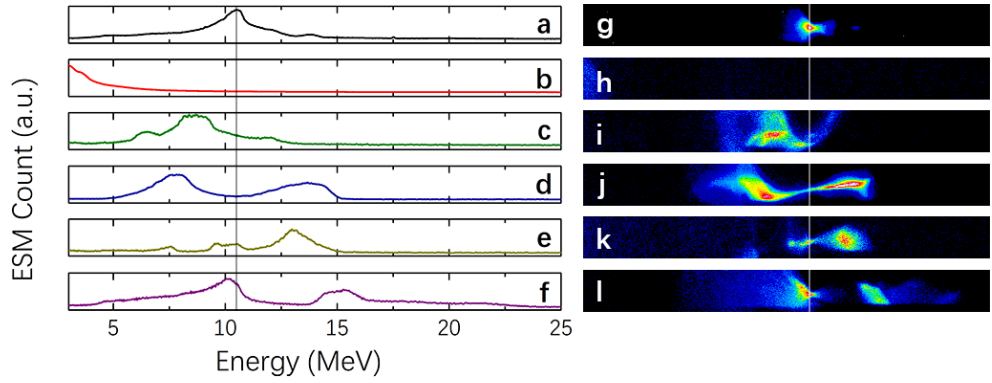

**Figure 3.** Output spectrum of staging acceleration. (a) Electron spectrum from the injector laser, focused by the solenoid. (b) Electron generated by the booster laser pulse only. (c)-(f) Electron spectrum with both injector and booster. (g)-(l) are the raw images taken by the ESM.

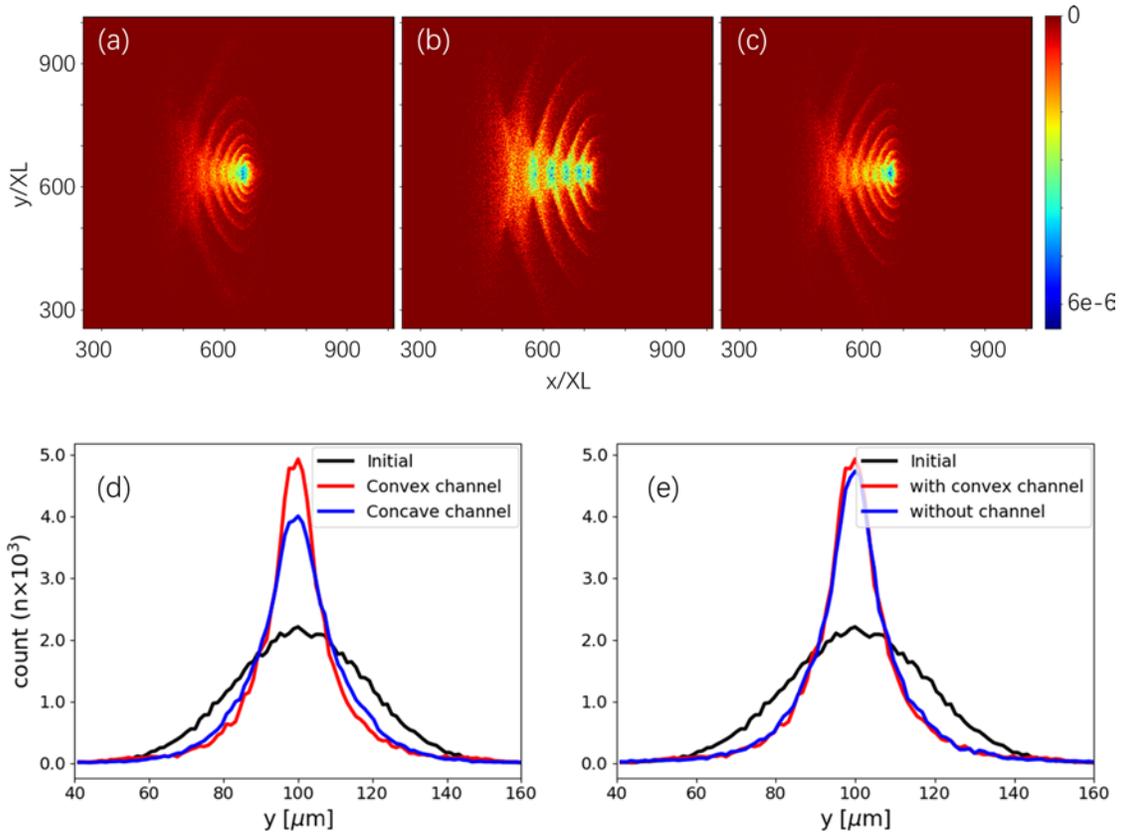

**Figure 4.** Self-focusing of an injector electron beam (Bennet-Budker effect) in low density plasma, $N_e=3\times10^{17}$ cm$^{-3}$. Fig. (a)-(c) give spatial distributions of bunch electrons after 30 ps propagation: (a) electron bunch, $E=10$ MeV, propagates in a concave plasma channel with $D=100$ μm and the depth 0.5 (b) electron bunch propagates in uniform plasma (c) electron bunch propagates in a convex channel with height 0.5. (d), (e) Transverse distribution of the electron bunch after 30 ps propagation in plasma with different shapes.

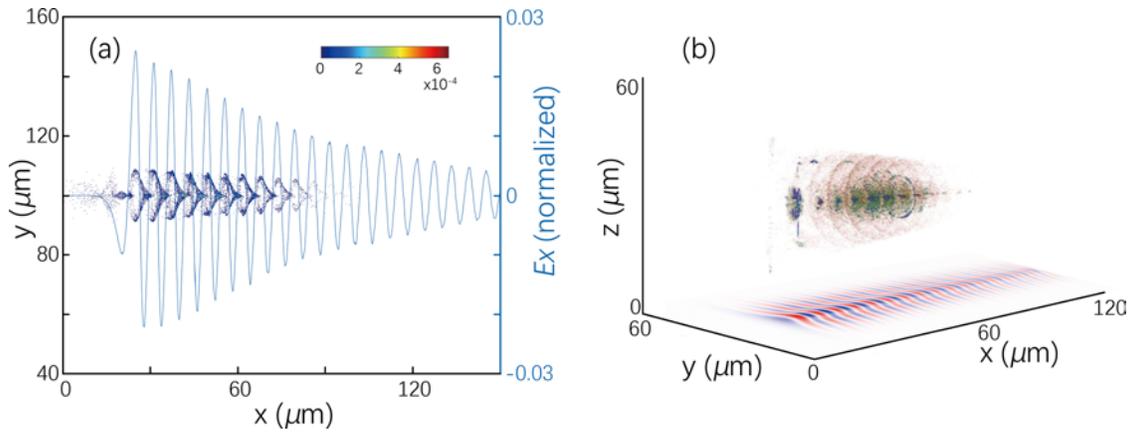

**Figure 5.** Spatial distribution of electron bunch (*E*=10 MeV, bunch length is 70 μm) modulated by the booster laser wake field and the wake field strength in uniform plasma; (a) 1D projection and (b) 3D for the bunch density with a 2D projection for the wake field.

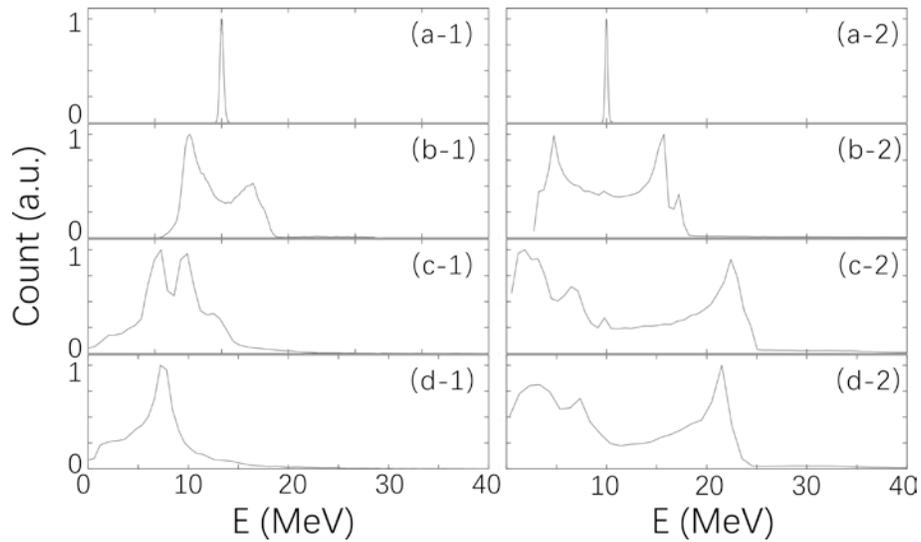

**Figure 6.** Evolution of electron beam energies during the interaction with the plasma waves in the second (booster) stage at different plasma density. Figure (a-1) to (d-1) shows electron energy evolution at plasma density of $3\times10^{19}$ cm$^{-3}$, and (a-2) to (d-2) show electron energy evolution at plasma density of $3\times10^{18}$ cm$^{-3}$ in the booster stage. (a-1,-2), (b-1,-2), (c-1,-2), and (d-1,-2) shows energy of the electron beam at 0.1 ps, 1.5 ps, 3.0 ps, and 4.5 ps, respectively.